\documentclass[pra,showpacs,twocolumn]{revtex4}
\usepackage{dcolumn}
\usepackage{bm}
\usepackage{amssymb}
\usepackage{amsmath}
\usepackage{amsfonts}
\usepackage[]{graphicx}
\begin{document}
\bibliographystyle{prsty}
\title{Near-field microscopy with a scanning nitrogen-vacancy color center in a diamond nanocrystalÿ: a brief review}
\author{A.~Drezet$^1$, Y. Sonnefraud$^1$, A.~Cuche$^{1,2}$, O.~Mollet$^{1,3}$, M.~Berthel$^1$, and S. Huant$^1$}
\address{($1$) Universit\'e Grenoble Alpes, Institut NEEL, F-38000 Grenoble, France and CNRS, Institut NEEL, F-38042 Grenoble, France.\\
         ($2$) CEMES CNRS UPR 8011, 29 rue J. Marvig, 31055 Toulouse Cedex 4, France.\\
         ($3$) Laboratoire de Photonique et de Nanostructures (CNRS-LPN), Route de Nozay, 91460 Marcoussis, France.}

\begin{abstract}
We review our recent developments of near-field scanning optical
microscopy (NSOM) that uses an active tip made of a single
fluorescent nanodiamond (ND) grafted onto the apex of a substrate
fiber tip. The ND hosting a limited number of nitrogen-vacancy (NV)
color centers, such a tip is a scanning quantum source of light. The
method for preparing the ND-based tips and their basic properties
are summarized. Then we discuss theoretically the concept of spatial
resolution that is achievable in this special NSOM configuration and find it to be only limited by the scan height over the imaged
system, in contrast with the standard aperture-tip NSOM whose
resolution depends critically on both the scan height and aperture
diameter. Finally, we describe a scheme we have introduced recently for high-resolution imaging of nanoplasmonic structures with ND-based tips that is
capable of approaching the ultimate resolution anticipated by
theory.
\end{abstract}

 \maketitle
\section{Motivation: beyond classical near-field microscopy}
Since its birth in the early 80's (Pohl et al., 1984), NSOM
(Courjon, 2003) became a versatile tool for optical imaging at very
high spatial resolution in the nanometer range (Novotny and Hecht,
2006). Yet one fundamental issue with NSOM is the optical resolution
offered by a given system. Standard systems based on
aperture-NSOM~(Pohl et al., 1984) with a hole at the apex of a
metal-coated conical tip are fundamentally limited by the size of
the optical aperture (Betzig and Chichester, 1993; Gersen et al.,
2001ÿ; Oberm\"{u}ller et al.; 1995a, Oberm\"{u}ller et al.; 1995b,
Drezet et al.; 2002, Drezet et al., 2004a). In order to improve the
optical resolution one could ideally use a
point-like emitting source.\\
Recently, inspired by the pioneer work by Michaelis \emph{et al.}
(Michaelis et al., 2000) who used a fluorescent single molecule at
low temperature as basis for a NSOM, we developed a high-resolution
NSOM tip that makes use of an NV center in a diamond nanocrystal as
a scanning point-like light source (Cuche et al., 2009a) (see
also~(Schr\"{o}der et al., 2011)). In this active tip the 20 nm
nanocrystal is glued \emph{in situ} to the apex of an etched optical
fiber probe. The NV center acts as a photostable (non blinking, non
bleaching) single-photon source working at room temperature (RT)
(Beveratos et al., 2002; Sonnefraud et al., 2008). As such, the
NV-center based tip proves to be superior to quantum-dot based
tips~(Chevalier et al., 2005), which suffer from insufficient
photostability (Sonnefraud et al., 2006), to insulating-nanoparticle
based tips, which, despite remarkable photostability, cannot reach
the single-photon emission rate (Cuche et al., 2009b), and to
tip-embedded light-emitting-diodes  (Hoshino et al., 2012), which
are quite involved to fabricate. Therefore, the ND-based NSOM probe
opens new avenues for microscopy and quantum optics in the
near-field regime.\\
In this paper we review our contribution to this field and discuss
the potentiality of such active-tip based NSOM in terms of spatial
resolution. Note that high-resolution apertureless NSOM (Zenhausern
et al., 1994; Bachelot et al., 1995) or NSOM based on specially
nanostructured passive tips (Mivelle et al., 2014; Eter et al.,
2014; Singh et al., 2014) are not covered by the present brief
review. In section 2 we describe in detail the fabrication process
of such tips and show how to characterize them. We emphasize in
particular the quantum properties of the NV emitters and show how to
characterize the photon emission statistics of the NV emitters at
the NSOM tip apex. In section 3 we show how to use such an active
tip for imaging and analyze its optical resolution. The theoretical
limit of the optical resolving power is discussed in section 4. and
the potentiality of the NV-based NSOM tip in the emerging field of
quantum plasmonics is shortly reviewed in section 5. Finally, we
present in section 6 some recent results concerning
nano-manipulation and displacement of NVs using a NSOM tip.

\section{NV center-based active tip}
Color centers in diamond~(Gruber et al., 1997), in particular NV centers, are very promising for the purpose of developing active NSOM. They are room-temperature single-photon
emitters~(Beveratos et al., 2001; Beveratos et al., 2002), their photostability is well
established~(Beveratos et al., 2001) and they
\begin{figure}[hbtp]
\begin{center}
\includegraphics[width=8.5cm]{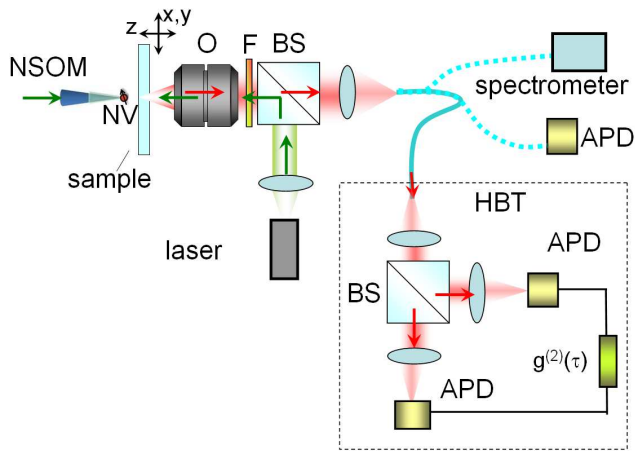}
\caption{Scheme of the optical setup used for tip functionalization
with a single fluorescent ND; (O= microscope objective, F= optical
filters and dichroic mirror, BS= beamsplitter, APD= avalanche
photodiode in the single-photon counting mode). The optical
excitation from an Ar$^+$-Kr$^+$ CW
laser is launched from the polymer-coated optical tip and the
NV-center fluorescence is collected by a high NA objective,
filtered, and injected into a multimode optical fiber. The latter
can be connected either to an APD, a spectrometer, or a HBT
correlator, which involves a 50/50 BS, two APDs and a
time-correlated single-photon counting module (for details see (Cuche et al., 2009)). The latter delivers the second-order
time-intensity correlation function $g^{(2)}(\tau)$: see text. Note that a rapid pre-selection of the NDs is usually made prior to the ND grafting: this is achieved by using the setup first in the confocal mode with the excitation being launched to the sample through the microscope objective O directly (green beam on the figure).}
\end{center}
\label{fig2}
\end{figure}are hosted by nanocrystals with steadily
decreasing sizes thanks to progress in material processing (Chang et
al., 2008; Sonnefraud et al., 2008; Boudou et al., 2009; Smith et
al., 2009). Early use of NV-center doped diamond nanocrystals in
NSOM active tips~(K\"{u}hn et al., 2001) was, however, limited by the size
of the hosting crystal, which was beyond the 50 nm range, so that
the promise of single NV-occupancy, i.e. single-photon emission, was
counterbalanced by size excess that prevents positioning with
nanometer accuracy. The recent spectacular reduction in size (Chang
et al., 2008; Sonnefraud et al., 2008; Boudou et al., 2009; Smith et
al., 2009) of fluorescent nanodiamonds (NDs), down to approximately
5 nm~(Smith et al., 2009), suggests that such limitation no longer
exists and that active optical tips made of an ultra-small (well
below 50
nm in size) ND with single NV-occupancy should be possible to achieve.\\
Our scanning single-photon sources are produced in a single
transmission NSOM (Sonnefraud et al. 2006; Sonnefraud et al., 2008)
environment. We successively use the optical tip for the imaging and
selection of the very ND to be grafted at the tip apex, for
controlled attachment of the latter, and subsequent NSOM imaging of
test surfaces.  A sketch of the optical setup is shown in Fig.~1.
After pre-selection of the NDs in the confocal geometry, the
NV-center emission is excited with the 488 or 515 nm line of an
Ar$^+$-Kr$^+$ CW laser that is injected by an uncoated optical tip
and is collected into a multimode optical fiber through a microscope
objective. The remaining excitation light is removed by means of a
dichroic mirror complemented either by a band-pass filter centered
at 607$\pm35$ nm for photon counting and imaging or a long-pass
filter ($>$ 532 nm) for spectra acquisition of the neutral and
negatively-charged NV centers~(Dumeige et al., 2004). The collection
fiber can be connected either to an avalanche photodiode (APD) for
imaging and optical control of the ND manipulation, to a
charge-coupled device attached to a spectrometer, or to a
Hanbury-Brown and Twiss (HBT) correlator for photon correlation
measurements. In the HBT module, a long-pass filter (750 nm) and a
diaphragm placed in front of each detector eliminate most of the
detrimental optical cross-talk.\\
The method that we have designed (Cuche et al., 2009a) to trap in a
controlled way a well-selected single ND at the optical tip apex is
as follows (see Fig.~2). The uncoated optical tip is covered with a
thin layer of poly-l-lysine, a polymer able to cover homogenously
the tip, including the apex (radius of curvature below 30 nm). In
addition, poly-l-lysine is positively charged. This facilitates
electrostatic attraction of the NDs, which bear negatively charged
carboxylic groups on their surface. This polymer-covered tip is
glued on one prong of a tuning-fork (Karrai and Grober, 1995)
\begin{figure}[h]
\begin{center}
\includegraphics[width=8.5cm]{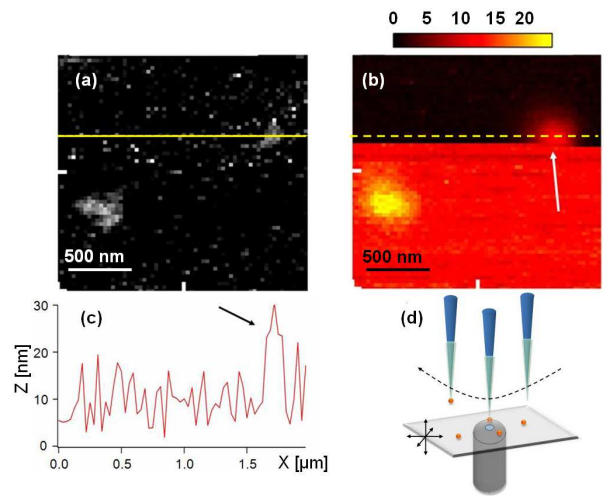}
\caption{ \textbf{(a)} Topographic and \textbf{(b)} fluorescence
NSOM images acquired simultaneously (scale bar in kilo-counts per
pixel). Images are recorded pixel by pixel by scanning the sample
under the tip from left to right and top to bottom (laser power at
the uncoated tip apex: 120 $\mu$W; integration time: 80 ms per
pixel; scanner speed: 1 $\mu$m s$^{-1}$; image sizes: 64 $\times$ 64
pixel$^2$). The line cut \textbf{(c)} (horizontal solid line in
\textbf{(b)}) gives the ND size (insert) at 20 nm in the present
case. The arrow in \textbf{(b)} marks the tip position where the
nanodiamond has been attached by the scanning tip. The principle of
the complete grafting experiment is sketched in \textbf{(d)} where
the NV centers, NSOM tip with polymer, scanning piezo-elements, and
microscope objective are all depicted. Images adapted from (Cuche et
al., 2009a)}

\end{center}
\label{fig2}
\end{figure} for shear-force feedback and mounted in the NSOM microscope.\\
The first step is to image the sample fluorescence to the far-field
by scanning the surface under the optical tip with a very large
tip-sample distance of 3 $\mu$m. This allows for selecting an
interesting area with isolated NDs. In a second step, the tip is
brought into the surface near-field by using shear-force regulation.
A near-field fluorescence image together with a shear-force
topography image are simultaneously recorded at a rather large
tip-sample distance of about 50 nm (usual cruise altitudes for NSOM
imaging are between 20 and 30 nm) in order to identify an isolated
small sized ND with a fluorescence level among the lowest-intensity
spots detected in the entire scanned area. This last point is taken
as a hint that this very ND presumably hosts a single color center.
This essential point can be checked in situ by photon-correlation
counting (Sonnefraud et al., 2008), before and
after grafting of the ND.\\
The next step is the ND attachment. This is accomplished "manually"
during scanning by strengthening the shear-force feedback (Karrai
and Grober, 1995) so as to approach the surface vertically to a
distance of around 30 nm with the optical tip facing the desired ND
(see Fig.~2(c) ). This shear-force strengthening is maintained for
typically two scanning lines and then released so as to bring the
now functionalized tip back to its initial altitude of 50 nm. Fig.~2
shows an example of a trapping event: It can be seen that the ND
trapping manifests itself as a sudden persistent increase in the
optical signal. This increase amounts to the emission level of the
ND prior to its attachment. Indeed, once the ND has been stuck at
the tip apex, the tip not only transmits the excitation laser light,
but also produces a background signal due to the attached ND,
irrespective of its position above the scanned surface. It is worth
noting in Fig.~2(b) that the shear-force feedback has been forced
(horizontal dashed line on the fluorescence image) after having made
sure that the ND height - 20 nm in the present case - has been
measured correctly in the topographic image (full line in
Fig.~2(a)). After attachment, the image acquisition is completed and
no additional ND is grafted to the tip due to the rather large
tip-to-surface distance of around 50 nm (see the sketch of principle
in Fig.~2(d)). This is the reason why the 40 nm in-height ND cluster
(possibly made of 3 NDs) seen in the lower left quarter in Fig.2~(b)
is not trapped by the scanning tip. It is worth noting that an
accidental fishing of an additional fluorescent ND would immediately
translate into an increase of fluorescence background emanating from
the tip: this provides us with a "safety procedure" ensuring that
such an
accidental fishing would be detected.\\
Now, the functionalized tip needs further optical characterization
since the attached ND was chosen from guesses that it would host a
single color-center. To check this important point, we carried out
photon-correlation measurements and spectrum acquisition of the
functionalized tip after having moved the tip far above the surface
(distance of 10 $\mu$m), laterally displaced the sample to a ND-free
region, and focused the collection objective onto the probe apex.\\
\indent The second-order time-intensity correlation function is
defined in quantum optics and in the stationary regime by
\begin{equation}
g^{(2)}(\tau)=\frac{\langle
E^{(-)}(t)E^{(-)}(t+\tau)E^{(+)}(t+\tau)E^{(+)}(t)\rangle}{\langle
E^{(-)}(t)E^{(+)}(t)\rangle^2 }
\end{equation}
where $E^{(\pm)}(t)$ are respectively the positive or negative
frequency part of the electric field operator associated with the
recorded photon stream. For a single-mode photon
field in particular, we can introduce the lowering $a(t)$ and raising
$a^{\dagger}(t)$ operators and get
\begin{equation} g^{(2)}(\tau)=\frac{\langle
a^\dagger(t)a^\dagger(t+\tau)a(t+\tau)a(t)\rangle}{\langle
a^\dagger(t)a(t)\rangle^2 }.
\end{equation}
In the case of a pure Fock state $|n\rangle$ corresponding to $n$
excitations or quanta in the mode, the previous expression takes the
simple form $g^{(2)}(\tau)= n(n-1)/n^2=1-1/n$.  This shows that for $n=1$, i.e. a single photon, $g^{(2)}(\tau)=0$. This result can be
easily understood if we interpret the $g^{(2)}$ function as a probability function:
\begin{equation}
g^{(2)}(\tau)=\frac{P(t+\tau|t)P(t)}{P(t)^2}=\frac{P(t+\tau|t)}{P(t)}
\end{equation} where $P(t)$ is the probability to detect one photon
at time $t$ (this probability is independent of $t$ in the
stationary regime) and $P(t+\tau|t)$ is the probability to detect a
photon at time $t+\tau$ conditioned on the detection of a photon at an earlier time $t$. For the $|n=1\rangle$ case we obviously have
$P(t+\tau|t)=0$ since a single photon can only be recorded
once.\\
The second-order correlation function is measured using the HBT
correlator shown in Fig.~1. The electric pulses generated by the two
single-photon counting detectors are fed into the ``start'' and
``stop'' inputs of an electronic counter/timer that both count the
number of pulses from each detector and also record the elapsed time
between subsequent pulses at the start and stop inputs. The
statistical accumulation of coincidence events gives a histogram of
the number of events recorded within a particular time interval.
\begin{figure}[hbtp]
\begin{center}
\includegraphics[width=8.5cm]{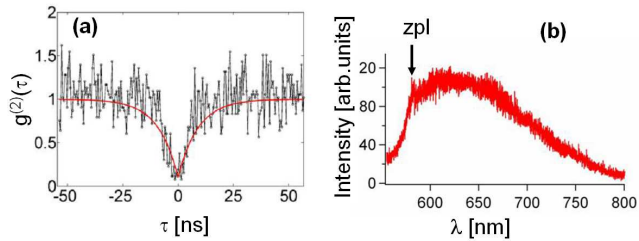}
\caption{\textbf{(a)} Normalized $g^{(2)}(\tau)$ function for the
functionalized tip giving evidence for single NV-center occupancy in
the functionalizing ND. The red curve is an exponential fit.
\textbf{(b)} Photoluminescence spectrum of a functionalized tip
(integration time: 180 s). The small peak at 575 nm (indicated by a
black arrow) is the zero-phonon line (zpl) of the neutral NV center.
Images adapted from (Cuche et al., 2009a)}
\end{center}
\label{fig3}
\end{figure} Fig.~3(a) shows $g^{(2)}(\tau)$ for the functionalized tip obtained after subtracting the random
coincidences caused by the background light (Sonnefraud et al.,
2008; Cuche et al., 2009a; Brouri et al., 2000). The correlation
function exhibits a clear anti-bunching dip at zero delay with
$g^{(2)}(0)$ dropping far below 0.5  (i.e. $g^{(2)}(0)\simeq 0.1$).
This unambiguously confirms that a single NV color-center, acting as
a single photon nano-source~(Cuche et al., 2009a), has been attached
at the tip apex. This NV center is an uncharged one as additionally
revealed by the optical spectrum of Fig.~3(b) which exhibits the
characteristic zero-phonon line of the neutral NV center at 575
nm~(Dumeige et al., 2004). We point out that the single photon
source does not generate pure Fock states. Indeed, in the considered
regime $g^{(2)}(\tau)$ takes the form

\begin{equation}
g_{\textrm{S}}^{(2)}(\tau)=1-\frac{1}{N}e^{-(\Gamma+r)\tau}
\end{equation} where $N$ is the number of incoherently excited NV
emitters, $\Gamma$ their typical spontaneous decay constant, and $r$
the excitation rate which, following Fermi rule, is proportional to
the laser excitation intensity. From Eq.~4 we deduce
$g^{(2)}(0)=1-1/N$, which indeed mimics the pure Fock states for
$n=N$. In the presence of an incoherent background taking into
account spurious fluorescence due to the sample and the tip itself,
Eq.~4 is modified to give
$g_{\textrm{S+B}}^{(2)}(\tau)=1-\frac{\rho^2}{N}e^{-(\Gamma+r)\tau}$,
where the coefficient $\rho=S/(S+B)$ defines the typical signal
intensity ($S$) to signal + background ($B$) light intensity ratio.
Fig.~3(a) takes into consideration this background light (up to the
spurious light coming from the ND itself) by comparing the recorded
fluorescence signal before and after the grafting event. The
correction $\rho$ allows us to represent directly the histogram
$g_{\textrm{S}}^{(2)}(\tau)$ and to compare it with Eq.~4. The fitting parameters confirm the value $N\simeq1$ (the small residual
value $g^{(2)}(0)\simeq 0.1$ is attributed to the background light
coming from the diamond crystal~(Cuche et al., 2009a) and from the finite
time resolution of the APDs and electronics). The value
$(\Gamma+r)^{-1}=9$ ns gives the red curve in Fig.~3(a). The pumping
rate is obtained using the condition $S=\eta r\Gamma/(r+\Gamma)$
with $S \simeq 9$ kHz$^{-1}$ the average photon rate on the APD, and
$\eta$ the total collection-detection efficiency of the whole
optical setup including the APDs. Considering the properties of the
different elements of the setup we estimate $\eta=1\%$. This leads
to $r^{-1}= 1$ $\mu$s and thus to the lifetime $\Gamma^{-1}\simeq 9
$ ns. This value concurs with those usually reported (i.e.
$\Gamma^{-1}\simeq 10-20$ ns) for the
considered nanodiamonds (Gruber et al., 1997; Beveratos et al., 2001; Brouri et al., 2000).\\
\indent We point out that our method is highly reproducible and
reliable. We have repeatedly functionalized tips on-demand with a
desired number of well-selected NDs in addition to the single ND
case described above. For example, as a demonstration of the
flexibility of our method, we have been able to successively glue five NDs with a tip, to free all of them at once by knocking the tip
on the surface, and to fish them back one by one~(Cuche et al.,
2010b). Additionnally, in contrast with previous attempts with
polymer-free tips which released the embarked nanocrystals within an
hour (Cuche et al., 2009c), the ND remains attached at the apex for
days so that the functionalized tip can not only be fully
characterized, but can also be used in subsequent experiments or
imaging.
\section{Imaging and spatial resolution} An interesting feature of the scanning single-photon
near-field source realized with the above technique is the spatial
resolution that it can potentially offer. In order to address this
issue we used it to image a test sample made of 250 nm wide and 40
nm thick chromium lines and parabola that have been lithographically
patterned on a fused silica cover slip (see Fig. 4(a)). The
collection light was spectrally restricted by proper filtering of
the NV emission and the scanning height was set at 20 - 30 nm. As
can be seen from Fig. 4(b), the optical image clearly reveals the
metallic structures as non-transmitting dark lines with a good
contrast even though it is recorded with the fluorescence light
emitted by a single NV center only. The distinctive bright
decorations seen on each side of the chromium structure are possibly
due to the finite optical reflectivity of chromium or to
modifications of the NV-center dynamical properties (e.g. change in
the excited state lifetime or local density of states) when the tip
probe approaches the nanostructure edges (Girard et al., 2005). Such
effects have been reported for ND emitters located
in the vicinity of metal nanostructures (Schietinger et al., 2009).\\
The chromium parabola in Fig. 4 has been patterned in such a way as
to offer a variable gap with the adjacent line. Our aim was to
infer a spatial resolution for our setup from its ability of
resolving two adjacent similar objects, in agreement with the basic
definition of a resolving power, rather than
\begin{figure}[hbtp]
\begin{center}
\includegraphics[width=8.5cm]{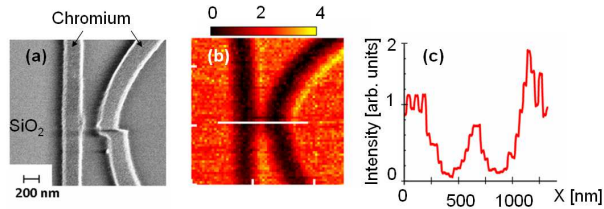}
\caption{\textbf{(a)} Scanning-electron micrograph of chromium
structures patterned on a fused silica cover slip.  \textbf{(b)}
Fluorescence NSOM image acquired simultaneously with the scanning
single-photon tip of Figs. 2 and 3 (optical power at 488 nm at the
uncoated tip apex: 120 $\mu$W, integration time: 100 ms, scan
height: h$\leq$ 30 nm, scanner speed: 1 $\mu$m s$^{-1}$, image size:
64 $\times$ 64 pixel$^2$, scale bar in kilo-counts per pixel). Here,
the collected light is restricted by optical filtering to the
emission band of the single NV center grafted on the optical tip.
\textbf{(c)} Cross-cut of the optical recorded signal along the
direction indicated by a white line in \textbf{(b)} (adapted from (Cuche et al., 2009a)) }
\end{center}
\end{figure}from the
lateral extension of the rise in the optical signal. In addition,
for this particular line-parabola doublet shown in Fig. 4., a
lithography failure brought incidentally the minimum gap to
approximately 120 nm (Fig.~4(a)). As seen in Fig. 4(b), this 120 nm
gap is resolved in the optical image. This indicates that the
spatial resolution is at least in this range, i.e, much better than
with the initial uncoated tip which offers resolutions limited to
about 400 nm (Sonnefraud et al., 2008). Furthermore, the cross
section of the optical intensity profile (see Fig.~4(c)) confirms
this finding and indicates that the resolution is at least in the
range 70- 150
nm. \\
Moreover, the near-field optical probe reported here acts as a
genuine scanning point-like dipole emitter. This contrasts with
metal-coated aperture tips which bear a polarization-dependent
annular charge density around the apex (Drezet et al., 2004b) that
limits the resolution (Drezet et al., 2004a; Drezet et al., 2004b;
Betzig and Chichester, 1993; Trautman et al., 1994; van Hulst et
al., 2000) and makes such tips mimic a dipolar behavior in the
far-field limit only (Oberm\"{u}ller et al., 1995a; Oberm\"{u}ller
et al.,1995b; Drezet et al., 2002), not in the near-field.
Point-like dipolar emitters do not exhibit such a split-field
distribution, so that their potential resolution should thus
ultimately depend on the scanning height $h$ only (see next
section). Interestingly, the ND-based active tip introduced here
could fully exploit this potentiality because the NV quantum emitter
is hosted by a matrix of a genuine nanometer extension. This
stresses the key role played by height control in future
developments. In the present proof-of-principle experiments, we set
a safe lower bound to the tip-surface distance at approximately
20-30 nm to avoid too strong friction forces applying to the tip
apex (Karrai and Grober, 1995), thereby preventing a too rapid
release of the 20-nm sized illuminating ND. We were then able to use
our functionalized tips for several days for image acquisition or
other measurements. This study clearly shows the potentiality of the
NV based NSOM for high resolution imaging. We shall now discuss on a
more theoretical basis the ultimate limit of such an active tip and
compare it with more usual NSOM aperture tips.
\section{Comparing a point-like emitter to an aperture NSOM tip}
In order to compare theoretically the spatial resolution offered by
a classical NSOM aperture tip with the one given by a point-like
dipole tip, we
\begin{figure}[h]
   \begin{center}
   \begin{tabular}{c}
   \includegraphics[width=8.5cm]{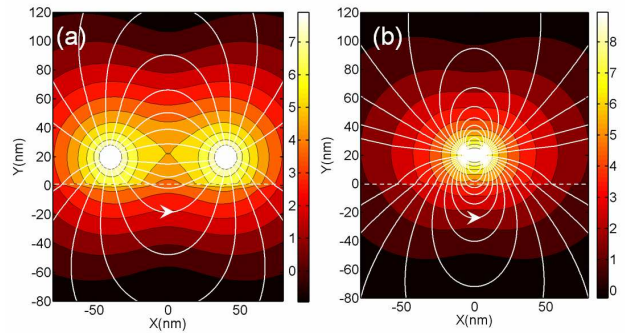}
   \end{tabular}
   \end{center}
\caption{Electric field generated by an aperture tip \textbf{(a)}
and a point-like tip  \textbf{(b)}. The field is calculated with the
tip in front of a glass substrate. The vertical distance $h$ between
both tips and the glass air interface is $h=20$ nm. The ring radius
is $a=40$ nm. For each panel the electric field lines and the
iso-density curves of the electric energy density $|\mathbf{E}|^2$
(in logarithmic scale) are calculated. The illumination wavelength
is $\lambda=600$ nm (adapted from (Drezet et al., 2011)) }
\end{figure}
first calculate the field generated by both probes. Fig.~5 shows a
comparison of the electric near-field generated by a ring-like
distribution (aperture radius $a=40$ nm, polarization along the $x$
axis) on the one hand (the method and calculation details are
described in (Drezet et al., 2004a; Drezet et al., 2011) and a
point-like dipolar source (dipole along the $x$ axis) on the other
hand. The comparison is made for a tip facing a glass substrate
(permittivity $\epsilon=2.25$) at a height of $h=20$ nm. In this
configuration the reflected and transmitted fields are calculated
with the image method, which is known to give consistent results in
the near-field zone (Drezet et al., 2011). The field generated by
the ring-like distribution contains both the electric and magnetic
contributions but since $ka$ and $kh$ are much smaller than unity we
checked that the effect of the magnetic as well as of the
propagating terms arising from the field propagator have negligible
effects (the same is true for the propagating terms
generated by the point-like dipole). However we keep all terms in the calculations for completeness.\\
In a second stage, we simulate an acquisition scan over an idealized
sample. The sample is made of either one or two point-like emitters
located at the glass-air interface. To simplify, we also assume that
the emitters are fluorescent particles emitting incoherently.
\begin{figure}[h]
\begin{center}
\begin{tabular}{c}
\includegraphics[width=8.5cm]{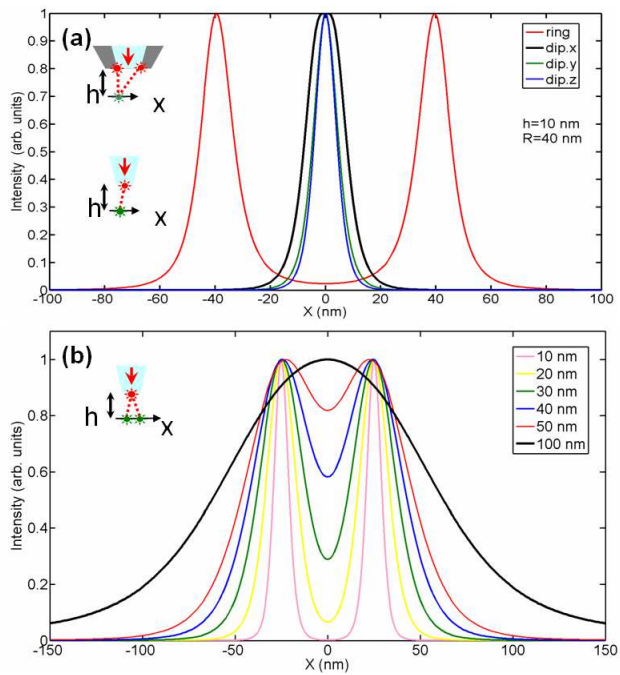}
\end{tabular}
\end{center}
\caption{\textbf{(a)} Simulations of the optical image obtained by
scanning a fluorescent isotropic emitter at a constant height $h=10$
nm below the NSOM tip in the ring-like and point-like
configurations, respectively (see inserts). The illumination
wavelength is $\lambda=600$ nm. The black curve is the theoretical
result obtained for a dipolar point-like source with a dipole
oriented along the $x$ direction. Similarly the green and blue
curves are the same images for a point-like dipole along the $y$ and
$z$ directions, respectively. These curves are compared with the
image obtained with an usual aperture NSOM, hole radius: 40 nm  (red
curve).\textbf{(b)} Simulations of the optical image obtained by
scanning two fluorescent isotropic emitters separated by a distance
$d=50$ nm (in the scan direction $x$) at a constant height $h$ below
the NSOM point-like probe (i.e., the NV active tip). The different
curves correspond to different $h$ going from 10 nm to 100 nm. The
probe dipole is along the vertical direction $z$ (adapted from
(Drezet et al., 2011)) }
\end{figure}The detection of the fluorescent
light through the substrate is done with a collection set-up (e.g.,
a microscope objective with high numerical aperture). More
precisely, we consider the photon absorption process by a nanosphere
located near $\mathbf{r}$ and containing an isotropic distribution
of fluorescent emitters. Following Glauber?s theory~(Glauber, 1963)
each emitter is excited by the field created by the tip with a
probability proportional to
$|\mathbf{E}(\mathbf{r},t)\cdot\mathbf{n}|^2$ where $\mathbf{n}$ is
the direction of the transition dipole associated with the
point-like fluorescent emitter located at $\mathbf{r}$. Here we
suppose a two-step process where the absorption is followed by a
fluorescence emission with probability $\eta(\omega')$  at the
emission frequency $\omega'$. After averaging over all possible
$\mathbf{n}$ we get a total fluorescence signal for the nanosphere
proportional to
$\eta(\omega')\cdot|\mathbf{\mathbf{E}}(\mathbf{r},t)|^2$ (Drezet et
al., 2004a; Drezet et al., 2004b). Note that in the case of the
single-photon tip this picture is very similar to the one used for
describing F\"{o}rster (or fluorescence) resonance energy transfer
(FRET) between two molecules~(Novotny and Hecht, 2006) (we however
neglect the back action of the molecular detectors on the dynamics
of the scanning dipole). Therefore the signal recorded at each tip
position is supposed to be proportional to the sum of the electric
energy density $|\mathbf{E}|^2$ at the location of the point-like
fluorescent
particles (Girard et al., 2005).\\
The collection efficiency of the NSOM microscope as used in (Cuche et al., 2009a) is defined by the properties of the high
numerical-aperture  objective and by the numerical aperture of the
multimode fiber which guides the collected light to the detector. We
estimate that 62-64\% of the $2\pi$ solid angle in which light is
emitted in the substrate is then collected by the optical setup. This justifies our assumption that essentially all polarisation components of
the electric field contribute to the optical signal
and therefore that this signal is proportional to $|\mathbf{E}|^2$.\\
Fig.~6(a) shows the variation of the optical signal during a scan
along $x$ for only one isotropic emitter on top of the substrate.
The comparison between both tips reveals important optical artifacts
with the ring-like NSOM tip due to the finite size of the ring and
to the high field intensity in the rim vicinity. These images can
easily be interpreted if we consider the fluorescent particle as a
test object moving in the near-field of the tips and scanning the
emission intensity profile in a plane at constant height $h$ above
the apex. The two peaks observed with the usual aperture NSOM are
well documented in the literature (Betzig and Chichester, 1993; Gersen et al., 2001; Drezet et al., 2004a) and are reminiscent of the high field available in the rim vicinity.
The point-like probe does not show such ``doubling'' of the imaged
structure and this eventually would lead to a simpler interpretation
of the optical images.\\
Having shown that the aperture tip leads to artifacts we consider
now the resolving power of the dipole tip. For this purpose we use two
scalar fluorescent emitters separated by a distance $d=50$ nm on the
glass substrate. The resolving power will be defined as the ability
to distinguish these objects with the NSOM tip. The aperture tip
diameter being larger than the gap $d$ we only analyze the
resolution of the dipole tip; for more details and comparisons see (Drezet et al., 2004a; Cuche et al., 2009a; Drezet et al., 2004b; Drezet et al., 2011). The results shown in
Fig.~6(b) for a dipole orientation along $z$ demonstrate that the
resolving power of this kind of microscope is ultimately limited by
the height $h$ only. Here, the system offers good resolution even
for $h=50$ nm, i.e. for $h\simeq d$. Only for $h\gtrsim d $ the
resolution will be dramatically degraded. These results are actually
very general for gaps $d$ much smaller than the wavelength since the
near-field of the probe, which is essentially wavelength
independent, dominates in this range.
\section{Quantum plasmonics: the resolution issue}
In this section, we discuss a method we have introduced recently that allows surpassing the resolution limit encountered in section 3 and approaching the theoretical limit discussed in the preceding section.\\
It is well known that a sub-wavelength object diffracts light into
evanescent and propagating waves. It is the evanescent part - the
so-called forbidden light - that carries information on the
sub-wavelength details of the object. This evanescent contribution
plays a key role in experiments targeted at imaging surface-plasmon
polaritons (SPPs), i.e. electron-photon hybrid states confined at
the boundary between a metal and an insulator. As such, SPPs are
strongly modified by local changes of their environment at the
nanoscale (Barnes et al., 2003; Novotny and Hecht, 2006) and it is
therefore critical to find efficient methods to probe the
interaction of SPPs with nanostructures. NV-based NSOM tips are
particularly adapted for this purpose since they are able to
interact locally with the plasmonic environment. Beside this imaging
facet, which can be understood in the context of classical Maxwell
electromagnetism, it is important to realize that we here enter the
realm of quantum optics, i.e., quantum plasmonics. This is because
NV centers are single-photon sources. Quantum plasmonics is an
emerging field with valuable prospects both on the fundamental
science and application agenda. One topic of large impact deals with
the coupling of single quantum emitters such as NDs with plasmonic
devices. In this context, early and more recent studies have shown
that fluorescent quantum emitters can efficiently couple to SPPs
when they are located in the vicinity of a metal structure
(Drexhage, 1974; Anger et al., 2006; Bharadwaj et al., 2007; Gerber et
al., 2007). Moreover, the possibility to generate individual SPPs
with single-photon sources opens the door to a wide range of studies
such as single-SPP mediated energy transfer (Chang et al., 2006;
Akimov et al., 2007; Fedutik et al., 2007; Wei et al., 2009),
locally-controlled enhanced fluorescence (Chang et al., 2006;
Schietinger et al., 2009), single-SPP
interferometry~(Kolesov et al., 2009), or SPP quantum interferences (Heeres et al., 2013; Fakonas et al., 2014; Di Martino et al., 2014).\\
Still, a fundamental understanding together with a tight control in
space, energy and polarization within this quantum regime is
essential to fully exploit these stimulating promises. Ideally, this
requires a deterministic control on the coupling of selected quantum
emitters to tailored plasmonic structures (Chang et al., 2006;
Girard et al., 2005; Liu et al., 2009). Recently, we made a decisive
step forward in this direction by demonstrating deterministic
launching of propagative quantum-SPPs at well-defined and freely
chosen positions into a nano-structured metal film by using NV-based
single photon tips (Cuche et al., 2010a; Mollet et al., 2011; Mollet
et al., 2012a; Mollet et al., 2012b). We have been able to
demonstrate that the $g^{(2)}(\tau)$ function of the NV sources is
fully conserved during the conversion of the evanescent light field
to
single SPPs and then back to radiative single photons~(Mollet et al., 2012a).\\
For this purpose we used leakage radiation microscopy (LRM) to probe
the propagation of SPPs along the metal film.
\begin{figure}[hbtp]
\centering
\includegraphics[width=8.5cm]{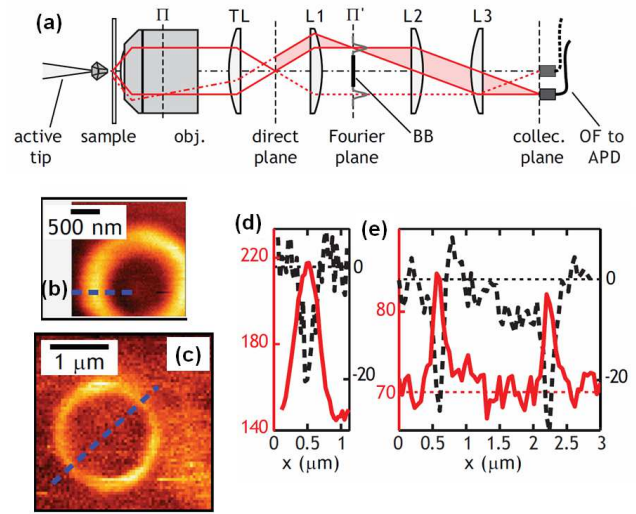}
\caption{\textbf{(a)} Layout of the plasmonic microscope: obj.= X100
oil immersion objective of effective numerical aperture $NA = 1.35$;
TL= tube lens; $L_1$, $L_2$ (removable), $L_3$= achromatic lenses;
$BB$= beam block; $OF$= multimode optical fiber; APD= avalanche
photodiode. $\Pi'$ is the back-focal plane of $L_1$. $\Pi$ is the
objective back-focal plane and is located inside the objective
itself. The $OF$-APD combination can be replaced by a camera (not
sketched) aligned with the optical axis for imaging. In this setup,
the tip is fixed and the sample is scanned in all three dimensions
with nanometer accuracy. The remaining excitation at 515 nm is
removed by an optical filter (not shown). A limited number of light
rays are indicated for clarity. \textbf{(b)} Direct-space image
obtained by scanning the slit under the ND tip. \textbf{(c)}
Reconstructed image obtained by mapping the intensity of the SPP
circle as function of the slit position under the tip. \textbf{(d)}
[respectively \textbf{(e)}] Cross sections along the blue dashed
lines in \textbf{(b)} [respectively \textbf{(c)}]. Left scales stand
for the optical signal levels, expressed in units of kHz, right
scales stand for the topography levels, expressed in nm (adapted
from Ref. (Mollet et al., 2012b))}
\end{figure}
Among the techniques used for imaging SPPs, LRM has emerged as a
powerful tool since it gives access in a rather straightforward way
to SPPs propagating along the interface between a dielectric and a
thin metal film (Hecht et al., 1996; Drezet et al., 2008; Hohenau et
al., 2011). In addition, as a far-field technique, LRM can analyze
SPP modes both in the direct and Fourier (momentum) spaces and it
has been successfully implemented in various plasmonic systems, both
at the classical (Baudrion et al., 2008; Stein et al., 2010; Wang et
al., 2011; Bharadwaj et al., 2011; Mollet et al., 2014)
and quantum levels (Cuche et al., 2010a; Mollet et al., 2011; Mollet et al., 2012a).\\
A sketch of LRM coupled to a NSOM is shown in Fig.~7(a). The
principle is based on the fact that for thin metal films confined
between air and glass, SPPs propagating at the air-metal interface
can leak through the film and evolve in the substrate as plane waves
emitted at a specific angle $\Theta_{LRM}$ (with respect to the
optical axis of the microscope). This angle is defined by
\begin{equation}
n_g\sin{\Theta_{LRM}}=n'_{SPP}(\omega)
\end{equation} where $n_g\simeq1.5$ is the optical index of glass and $n'_{SPP}(\omega)$ is the real part of the SPP in-plane index at the optical wavelength
$\lambda=2c\pi/\omega$~(Barnes et al., 2003; Drezet et al., 2008).
It corresponds to a value larger than the critical angle in glass
$\Theta_c=\arcsin{(1/n_g)}$ ($\Theta_c\simeq 43.2^\circ$ in fused
silica of optical index $n_g \simeq 1.46$). Therefore, leaky SPPs
contribute to the `forbidden' light sector (Novotny and Hecht,
2006). In our setup (Mollet et al., 2012b) we use several lenses for imaging
the SPP propagation either in the direct space (conjugated with the
air-metal interface of the sample) or in the Fourier space (see
Fig.~7(a)). LRM can be combined with a NSOM and Fourier filtering
techniques to image SPPs only (Cuche et al., 2010a; Mollet et al.,
2011; Mollet et al., 2012a; Mollet et al., 2012b). Our imaging
method consists in reconstructing optical images solely from the
plasmonic `forbidden' light collected in the Fourier space. It is
demonstrated below by using a point-like ND-based tip that
illuminates a thin gold film patterned with a sub-wavelength annular
slit.\\
The ND is illuminated through the tip with a $\lambda_{exc.}= 515$
nm laser light. As shown previously (Cuche et al., 2010a; Mollet et
al., 2011; Mollet et al., 2012a; Mollet et al., 2012b), the
red-orange near-field fluorescence of the NVs launches SPPs into the
gold film, whereas the green laser excitation cannot do that because
of strong interband absorption in gold in this wavelength range:
this is an additional motivation for using a ND tip. This
arrangement generates a range of undesired cross-excited
nonplasmonic light such as gold fluorescence in addition to the
useful plasmonic signal: these photons are filtered using a mask
located in the Fourier space ($BB$ in the plane $\Pi'$ of
Fig.~7(a)). In the quantum regime, where only a few photons couple
to SPPs, it is crucial to eliminate this nonplasmonic spurious
light. The annular slit has a $\simeq$ 120 nm rim thickness and
$\simeq$ 1.5 $\mu$m inner diameter. It is patterned by focused-ion
beam (FIB) milling in a 30 nm thick gold film. The imaging results
are shown in Fig.~7(b-e). It is clear that the reconstructed image
in Fig.~7(c) is much better resolved than the direct space image
depicted in Fig.~7(b). This is because in the first case the image
was recorded by removing all non plasmonic signals~(Mollet et al.,
2012a; Mollet et al., 2012b), whereas in the second case the image
is recorded by collecting photons in the direct space irrespectively
of their in-plane momentum. As a matter of fact, the sharpness of
the reconstructed image competes with that of the simultaneously
acquired topographic image~(Mollet et al., 2012b) (not shown here).
This is confirmed by the cross sections shown in Figs.~7(d) and (e).
In Fig.~7(e), the full width at half maximum of the optical signal
is around 130 nm, to compare with 100 nm in the corresponding
topographic cross section (Mollet et al., 2012b), whereas in
Fig.~7(d) it is as large as 230 nm, compared with 120 nm in the
topography. Therefore, reconstructed images from the
Fourier-filtered signals made only of high spatial frequencies, i.e.
those due to SPPs leaking in the silica substrate, exhibit a four
times enhanced spatial resolution of $\simeq$ $130-100=30$ nm
compared to $\simeq$ $230-120=110$ nm obtained in the direct space.
This clearly shows the advantage of the reconstruction method
concerning resolution. It is worthwhile to note that this 30 nm
resolution fits well with the ND size and with the typical distance
between tip and surface (Karrai and Grober, 1995) during scanning.
Therefore, this method offers a resolution that approaches the
ultimate spatial resolution achievable with a point-like optical tip
as discussed in section 4 (Drezet et al., 2011).
\section{Nano-manipulation of NV centers using a NSOM tip}
The possibility to translate precisely a quantum emitter in a
structured environment is of tremendous importance for numerous
applications, see e.g. (Cuche et al., 2010a; Beams et al., 2013;
Schell et al., 2014; Geiselmann et al., 2013; Tisler et al., 2103;
Rondin et al., 2014). One application is quantum plasmonics. Recent
studies focused on the possibility to locate precisely a quantum
emitter near an antenna in order to boost or control its
fluorescence/luminescence properties. This is in particular the case
for works with an atomic force microscope (AFM) tip used to control
precisely the distance between an NV and some gold nanoparticles
(Schietinger et al.; 2009) in order to reduce the lifetime of the
emitter (an application which could be important for 2D quantum
nano-technology). The same group used different approaches based on
AFM methods in order to manipulate nano-objects: (i) pushing or (ii)
`fishing'. The first method could be compared to a `nano-golf'
method in which the AFM tip pushes in a gentle way the nanoparticle
(Schietinger et al.; 2009) . \begin{figure}[hbtp] \centering
\includegraphics[width=8.5cm]{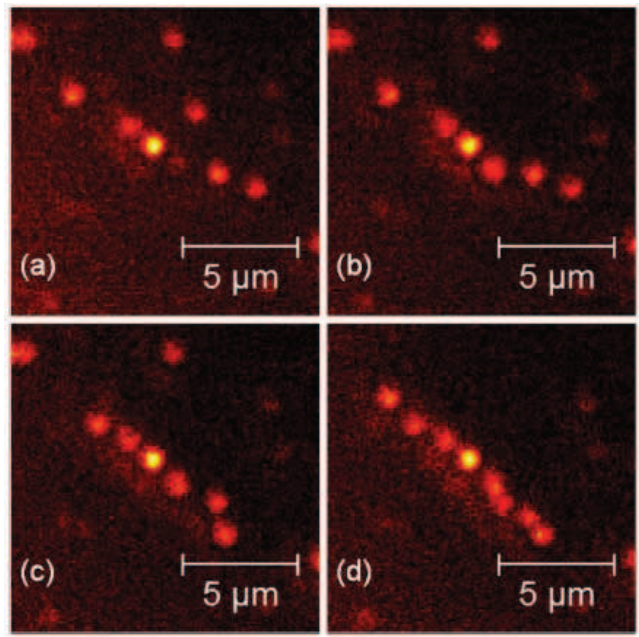}
\caption{\textbf{(a-d)} An illustration demonstrating how a NSOM tip
can be used to align 8 fluorescent diamonds (80 nm diameter) on a
glass substrate.}
\end{figure}The second method is more demanding and
requires to be able to pick up a nanoparticle with a tip (using e.g.
the NSOM procedure presented in this manuscript) and then to relax
the particle in a controllable way in a given environment (Schell et
al., 2011). The clear advantage of the second method is that we can
translate the nano-objets over large distances, which is very
relevant if these objects are
first deposited in a region free of nanostructures.\\
Due to
their interest for applications, we have also investigated the above two methods using NV-based active NSOM tips.
Here, we present an example of both approaches. Fig.~8 shows an
example of the first procedure. The aim is to translate NV centers
contained in 80 nm diameter NDs by using a bare NSOM
tip to push and align up to 8 fluorescent NDs.
Fig.~8(a-d) show different stages of the experiment involving
different number of diamonds. Such an alignment procedure is
expected to be of interest in a periodical (plasmonic) system or
near an antenna to study the coupling between plasmonic and photonic
modes. Fig.~9 shows an example of the second approach (see also (Cuche et al., 2010b) for other demonstrations).\begin{figure}[hbtp]
\centering
\includegraphics[width=4cm]{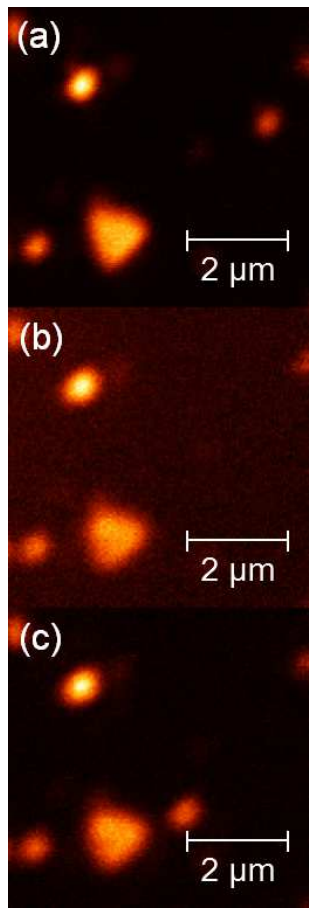}
\caption{\textbf{(a-c)} An illustration showing how a NSOM tip
can be used to first graft and subsequently release a single fluorescent ND (80
nm diameter) on a glass substrate near a plasmonic triangular
nano-plate. In (a), the ND is located in the upper right corner of the image. In (b), it has been attached onto the NSOM tip and is no longer visible on the image. In (c), it has been re-deposited on the substrate near the gold triangle in the lower part of the image.}
\end{figure} Here, the tip with
the polymer coating is used to pick up a single
fluorescent ND. This diamond is subsequently released at a
different position near a colloidal gold nano-prism (10 nm thick and
approximately 500 nm side lengths). The method was repeated several
times to confirm its reproducibility. The aim of the approach would be
to control the lifetime of the emitter near the edges of the
triangle.
\section{Conclusion}
In this article, we have reviewed our contributions to the
development of active-tip based NSOM, with a special emphasis on
active tips made of a fluorescent ND. In the case of a single NV
occupancy in the ND, such a tip forms a genuine scanning
single-photon tip. We have discussed the ultimate resolution
achieved by such a tip in optical imaging and shown that it is
limited by the scan height only, in contrast with standard aperture
tips, which are resolution-limited by the aperture size as well. By
applying a ND-based tip in SPP launching into a nanostructured gold
film and reconstructing images from the Fourier-space SPP signal, we
have shown that the scanning plasmonic microscopy achieved this way
is able to approach this ultimate resolution. Finally, we
demonstrated that the active-based NSOM method is also a versatile
tool for moving and positioning precisely NVs in a 2D environment.
This represents a promising avenue for future nano-manipulation of
quantum emitters in a plasmonic system.\\

AD, AC, OM, MB and SH wish to dedicate this review article to their
co-author and colleague Yannick Sonnefraud who passed away in
September 2014. Yannick initiated this research in 2008 (Sonnefraud et al., 2008).

\end{document}